\documentclass[a4paper,11pt]{article}
\usepackage{pos}
\usepackage{subfigure,wrapfig}

\title{Contribution to Galactic cosmic rays from young stellar clusters}
 \ShortTitle{Cosmic rays from young stellar clusters}

\author*[a]{G. Morlino}
\author[a]{S. Menchiari}
\author[a]{E. Amato}
\author[a]{N. Bucciantini}

\affiliation[a]{INAF/Osservatorio Astrofisico di Arcetri, Largo E. Fermi, 5 - 50125 Firenze, Italy}

\emailAdd{giovanni.morlinio@inaf.it}

\abstract{The origin of Galactic cosmic rays (CR) is still a matter of debate. Diffusive shock acceleration (DSA) applied to supernova remnant (SNR) shocks provides the most reliable explanation. However, within the current understanding of DSA several issues remain unsolved, like the CR maximum energy, the chemical composition and the transition region between Galactic and extra-Galactic CRs. These issues motivate the search for other possible Galactic sources. Recently, several young stellar clusters (YSC) have been detected in gamma rays, suggesting that such objects could be powerful sources of Galactic CRs. The energy input could come from winds of massive stars hosted in the clusters which is a function of the cluster total mass and initial mass function (IMF) of stars.
In this work we evaluate the total CR flux produced by a synthetic population of YSCs assuming that the CR acceleration occurs at the termination shock of the collective wind resulting from the sum of cluster’s stellar winds. We show that the  spectrum produced by YSC can significantly contribute to energies $\gtrsim 100$\,TeV if the diffusion inside the wind-blown bubble is Bohm-like and the spectral slope is harder than the one produced by SNRs.}

\ConferenceLogo{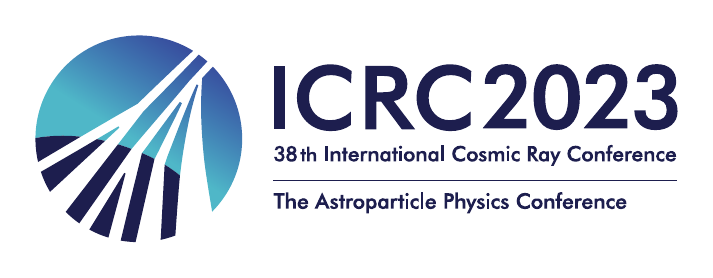}

\FullConference{%
38th International Cosmic Ray Conference (ICRC2023)\\
  26 July - 3 August, 2023\\
  Nagoya, Japan}

\begin{document}
\maketitle

\section{Introduction}
\label{sec:into}
The origin of Galactic Cosmic Rays (GCRs) in the knee region is still debated in the community. Recently, young and massive stellar clusters (YMSCs) have been suggested as alternative candidate sources to supernova remnants (SNRs). The energy input could come from the cluster stellar winds which provide, during a lifetime of several million years, an energy comparable with respect to SNR outputs \cite{Cesarsky-Montmerle:1983}. In addition, YMSCs could offer favorable conditions for particle confinement. The stellar winds, in fact, generate a wind-blown bubble around the cluster with a typical size of tens of pc, where the magnetic turbulence may be enhanced with respect to the Galactic interstellar medium (ISM), increasing the particle diffusion time and, consequently, the possibility to achieve very high energies.
Besides energetic considerations, acceleration of particles from the wind of massive stars appears to be a necessary component to explain the ${}^{22}$Ne/${}^{20}$Ne anomaly in CR composition \cite{cassePaul}: the enhancement observed with respect to Solar abundances  requires acceleration from a carbon-enriched medium rather than from the standard ISM at a few percent level. The relative contribution of these different source populations (SNRs and YMSC) to the observed CRs is yet to be clarified, in particular in the region across the \emph{knee} around few PeVs.  

In recent years, diffuse gamma-ray emission has been detected in coincidence with many young massive stellar clusters (YMSC) by several gamma-ray facilities, like Fermi-LAT, H.E.S.S. \cite{aharonian} and LHAASO. These findings strongly supporting the idea that some acceleration mechanism is taking place there. Detailed morphological and spectral analysis suggest a continuous injection of possible hadronic origin \cite{MenchiariPhD:2023,Blasi-Morlino:2023}. The Cygnus cocoon has even been observed at the highest energies ever probed by gamma-ray astronomy: a 1.4~PeV photon was detected from LHAASO \cite{lhaaso}, strengthen the idea that  stellar clusters may act as PeVatrons.

The exact location where particle acceleration takes place in YMSCs is still unclear. For compact and young systems, the so-called wind termination shock (WTS) \cite{Morlino+2019} is expected to be strong enough to enable particle acceleration at such high energies \cite{gupta}. Alternatively, stochastic acceleration might be driven by the highly turbulent environment of the cluster, particularly in its core \cite{badmaev}, further amplified once SN explosions start to occur \cite{vieu1}. 
In this work we want to estimate the contribution of YMSCs to the Galactic CR flux before supernova start to explode. Hence, we will consider only acceleration at the cluster WTS. For each SC, we build a stellar population, which properties in terms of wind speed and mass loss rate are used to  determine the size of the wind-blown bubbles, as described in \S\ref{sec:wind}. Then, we will integrate the WTS contribution over the entire population of Galactic SCs. However, the SC population is reasonably well know only within $\sim 2$\,kpc from the Sun and, to overcame this lack of information, we will build a synthetic population of SC based on properties of local clusters, as explained in \S\ref{sec:clusters}. Finally, in \S\ref{sec:acceleration} we summarize the acceleration model of \cite{Morlino+2019} and we discuss the results in \S\ref{sec:results}. 
The reader is also referred to a companion work presented in the same conference \cite{Menchiari-ICRC:2023a} where the same approach is used to estimate the SC contribution to the Galactic diffuse gamma-ray emission.

\section{Properties of stellar winds and wind-blown bubbles}
\label{sec:wind}
To describe the bubble structure around a SC, for each star we need two quantities, the wind velocity and the mass loss rate and the star age. In the following we will only deal with main sequence stars, neglecting the contribution from the stellar final stages, like Wolf-Rayet, which, however, may contribute up to $\sim 30\%$ of the wind power in a SC \cite{Cesarsky-Montmerle:1983}. 
In the stellar wind theory, the wind velocity is generally written as \cite{Kudritzki-Puls:2000} 
\begin{equation} \label{Vwind_star}
 v_{\rm w,\star} = C(T_{\rm eff}) v_{\rm esc} 
 = C(T_{\rm eff}) \left[ 2 g R_{\star}\, (1-\Gamma) \right]^{1/2}
\end{equation}
where $v_{\rm esc}$ is the escape speed form the star, $g$ is the surface gravity and $R_\star$ the stellar radius. The factor $\Gamma=L_{\star}/L_{\rm Edd}$ takes into account the reducing effect of Thomson scattering on the gravitational potential. The wind velocity is in general larger than  $v_{\rm esc}$ due to the radiation pressure from the star. Such an effect is accounted for by the function $C$ which depends on the effective surface temperature of the star which is, in turn, estimated using the Boltzmann law: $T_{\rm eff}=\left [\frac{L_\star}{4 \pi R_{\star}^2 \sigma_b} \right ]^{1/4}$, $\sigma_b$ being the Boltzmann constant. $C$ ranges from 1 for $T_{\rm eff}<10^4$\,K up to 2.65 for $T_{\rm eff} > 2 \times 10^4$\,K. In this range we assume a linear increase with $T_{\rm eff}$. 

The stellar mass loss rate is a rather difficult quantity to constraint from observations. Stellar evolution codes provide results which, however, depends on several quantities like the metallicity and the the stellar rotation. For the sake of simplicity, here we use the approximate model by \cite{Nieuwenhuijzen-deJager:1990} where the mass loss rate depends only on the stellar luminosity, mass and radius, and reads \cite[for a comprehensive discussion see also][]{MenchiariPhD:2023}:
\begin{equation} \label{Mdot_star}
 \dot{M}_{\star} = 9.55 \times 10^{-15} (L_{\star}/L_{\odot})^{1.24} (M_{\star}/M_{\odot})^{0.16} (R_{\star}/R_{\odot})^{0.81} \, {\rm M_{\odot} \, yr^{-1}} \,.
\end{equation}
The stellar luminosity is only a function of its mass and it is taken from \cite{MenchiariPhD:2023} and consists of a smoothed broken power law mixing two different empirical mass-luminosity relations. The relation between stellar radius and mass is given to first approximation by $R_{\star}/R_{\odot} = 0.85 (M_{\star}/M_{\odot})^{0.67}$ \cite{Demircan-Kahraman:1991}. We notice that the uncertainty in the mass-ratio relation translates into an uncertainty of only $\sim 15\%$ in the final cluster's luminosity. For the purpose of this work, the properties of stellar winds can be considered almost stationary during the main sequence lifetime, which last 
\begin{equation} \label{Age_star}
 \log{\left(T_{\rm age}/\rm yr\right)} = 6.43 + 0.825 \, \left[\log{(M_{\star}/120 \rm M_{\odot})}\right]^2 \,.
\end{equation}
After such a time, we assume that the stellar wind do not contribute anymore to the SC wind. We stress again that the subsequent explosion of SNe from massive stars is neglected.


For the initial mass function (IMF) of star inside a cluster, we adopt the distribution from \cite{Kroupa:2001} which is a broken power-law in several mass range, $\xi_{\star}= A_i(M_{\rm cl}) M_{\star}^{k_i}$, which reduces to a Salpeter IMF for $M > M_{\odot}$ with $k_i=2.35$. The IMF for each cluster is normalized to give the SC total mass, i.e. $\int_{M_{\star,\min}}^{M_{\star,\max}} M \xi_{\star}(M_{\star},M_{\rm cl}) dM = M_{\rm c}$.
The minimum and maximum stellar mass are assumed to be $M_{\star,\min}=0.08 M_{\odot}$ which is related to the minimum theoretical mass to support significant nuclear burning and $M_{\star,\max}=150$ M$_\odot$ that is the maximum observed stellar mass.

Once the stellar distribution is fixed, the properties of the SC collective wind can be estimated from the mass and momentum flux conservation, integrating over all stellar winds. Hence, the final SC mass loss rate and wind speed are:
\begin{eqnarray}
 \dot{M}_{\rm c}(M_{\rm c}) = \int_{M_{\star,\min}}^{M_{\star,\max}} \dot{M}_{\star} \, 
 \xi_{\star}(M_{\star},M_{\rm c}) \, dM_{\star} \\
 v_{\rm w, c} = \frac{1}{\dot{M}_{\rm c}}
   \int_{M_{\star,\min}}^{M_{\star,\max}} \dot{M}_{\star} v_{\rm w,\star} \, \xi_{\star}(M_{\star},M_{\rm c}) \, dM_{\star} 
\end{eqnarray}
The bubble structure is described using the classical solution for adiabatic expansion \cite{Weaver+1977}. Defining the cluster age $t$ and the cluster luminosity $L_{\rm w,c}=\frac{1}{2}\dot{M}_{\rm c} v_{\rm w,c} ^2$, as well as the ISM mass density $\rho_0$, the position of the TS is located at
\begin{equation}
\label{eqn:Rs}
R_{\rm s}(t) = 48.6 \, \left(\frac{\rho_0}{\mathrm{cm}^{-3}}\right)^{-0.3} \left(\frac{\dot{M}_{\rm c}}{10^{-4}\mathrm{M}_\odot\mathrm{yr}^{-1}}\right)^{0.3} \left(\frac{v_{\rm w,c}}{1000\,\mathrm{km\,s}^{-1}}\right)^{0.1} \left(\frac{t}{10\,\mathrm{Myr}}\right)^{0.4} \,\mathrm{pc}
\end{equation}
while the bubble radius is
\begin{equation}
\label{eqn:Rb}
R_{\rm b}(t) = 174 \, \left(\frac{\rho_0}{\mathrm{cm}^{-3}}\right)^{-0.2} \left(\frac{L_{\rm w,c}}{10^{37}\,\mathrm{erg\,s}^{-1}}\right)^{0.2} \left(\frac{t}{10\,\mathrm{Myr}}\right)^{0.6} \mathrm{pc} \, .
\end{equation}
Because stellar clusters born from giant molecular clouds, the local ISM density where their winds expand is usually denser than the average Galactic ISM. Here we assume a reference value of $\rho_0=10$~protons/cm$^3$ identical for all SCs.

\section{Stellar cluster distribution}  
\label{sec:clusters}
\begin{figure}
\centering
\subfigure[\label{fig:mdot}]{\includegraphics[width=0.49\textwidth]{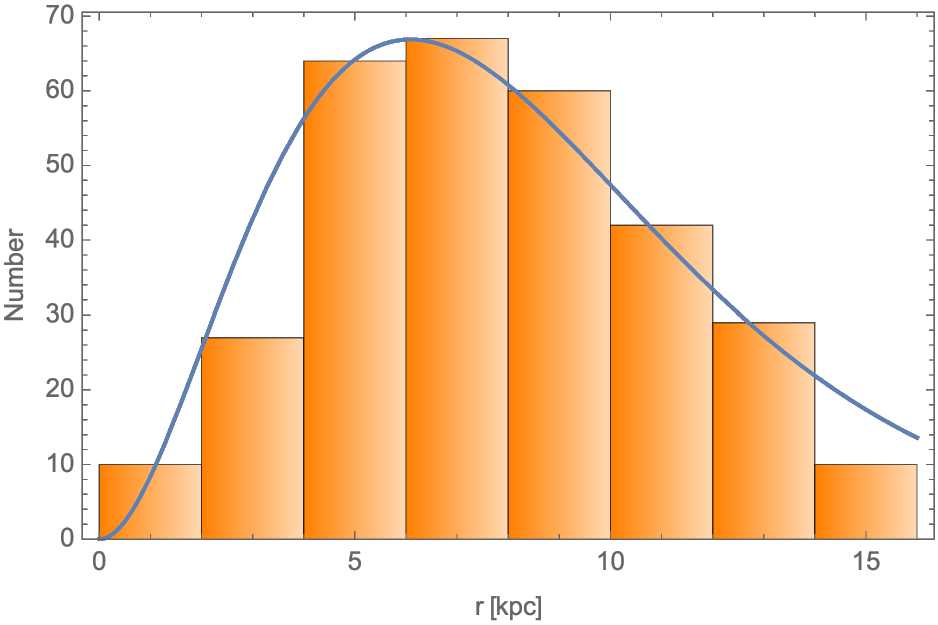}}
\subfigure[\label{fig:lum}]{\includegraphics[width=0.45\textwidth]{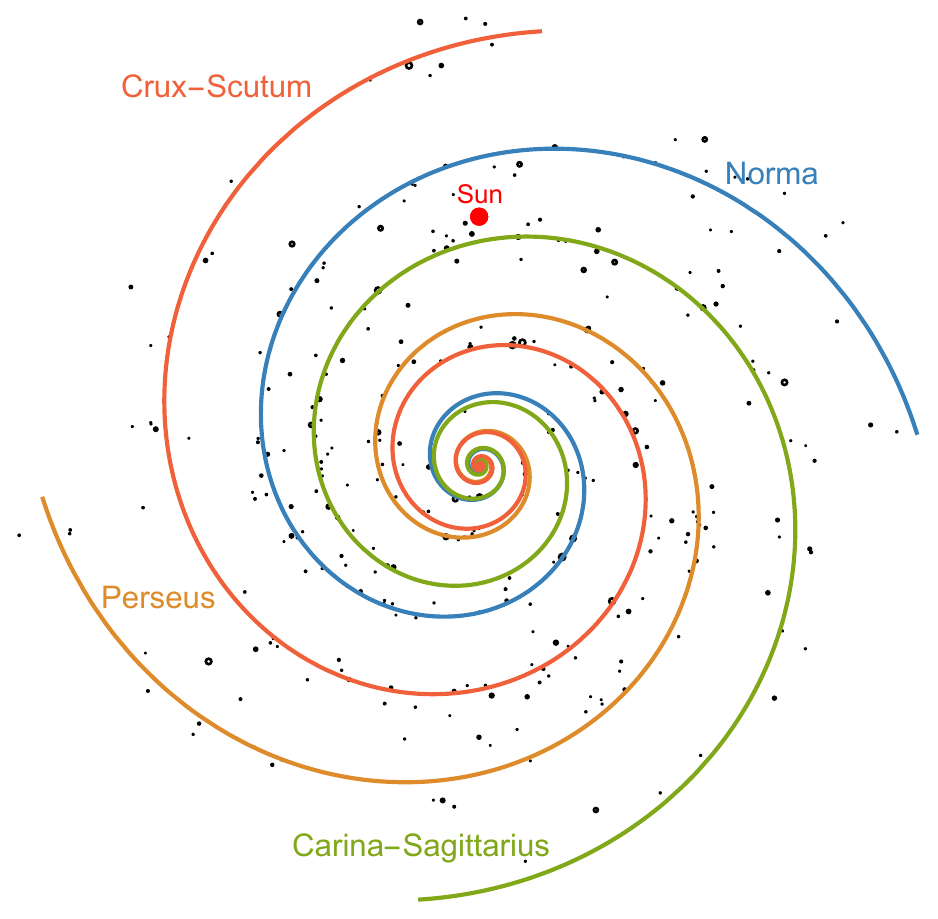}}
\caption{Distribution of a single realization of stellar cluster population with age $< 3\, \rm Myr$, as a function of Galactocentric radius (a) and in the Galactic plane (b). The solid line in (a) show the pulsar distribution from Eq.\eqref{eq:pulsar_dist} for comparison.}
\label{fig:clusters}
\end{figure}

The SC distribution in mass, time and position in the Galaxy is defined as 
\begin{equation}
\label{eqn:SC_dist}
 \xi_{\rm c}(M_{\rm c},T,\vec{r}) = \frac{dN}{dT dM_{\rm c} d\Sigma}\, ,
\end{equation}
such that $\xi_{\rm c} dT dM_{\rm c}$ is the number of SC with initial masses $[M_{\rm c}, M_{\rm c} + dM_{\rm c}]$ formed per unit surface $d\Sigma$ of the Galactic disk, at the position $\vec{r}$, in the time interval $[T, T + dT]$. Following  \cite{Piskunov+2018} we assume that the distribution is factorized in time and mass. Moreover, we also assume that the distribution can be factorized in space such that the cluster initial mass function (CIMF) depends only by the distance $r$ from the Galactic Centre through a normalization factor. Hence, we can write
\begin{equation}
\label{eq:SC_dist2}
 \xi_{\rm c}(M_{\rm c},T,\vec{r}) = \psi(T) f_{\rm c}(M_{\rm c}) \rho_{\rm c}(r) 
\end{equation}
where $\psi$ is the SC formation rate (CFR), $f_{\rm c}(M_{\rm c})$ is the CIMF and $\rho_{\rm c}(r)$ is the cluster radial distribution, normalized to be unity at the Sun location, $r_{\odot}= 8.5$\,kpc.
To get the present distribution of SC we should integrate in time Eq.\eqref{eq:SC_dist2}, however, here we are interested in describing only the young population of SC, with an age $\lesssim 10$\,Myr, because for larger ages the wind power drops to negligible values and the production of CRs becomes irrelevant. \cite{Piskunov+2018} showed that the present SC distribution in the solar neighborhood is compatible with a formation rate roughly constant during the last $\sim 50$\,Myr. Hence the CFR can be assumed constant. Its value can be derived from the work by \cite{Lamers-Gieles:2006} who obtained a surface star formation rate in the solar neighbourhood of $350 \, \rm M_{\odot} \, Myr^{-1} \, kpc^2$ for clusters' mass between $100\, \rm M_{\odot}$ and $3\times 10^4\, \rm M_{\odot}$. This corresponds to an average CFR of $\bar{\psi} = 0.63\, \rm kpc^{-2} \, Myr^{-1}$.
For the CIMF, we follow \cite{Piskunov+2018} which derived for the SC population in the solar neighborhood the following broken power-law:
\begin{equation}
\label{eq:CIMF}
 f_{\rm c}(M_{\rm c}) = \,
 \begin{cases}
     k_1 \, M_{\rm c}^{-1.63} \hspace{0.5cm} {\rm for} \hspace{0.5cm} M_{\rm c,\min} <M_{\rm c} < M_{\rm c}^* \\
     k_2 \, M_{\rm c}^{-1.24} \hspace{0.5cm} {\rm for} \hspace{0.5cm} M_{\rm c}^* <M_{\rm c} < M_{\rm c,\max}
 \end{cases}
\end{equation}
where $M_{\rm c,\min}= 2.5 \,\rm M_{\odot}$, $M_{\rm c,\min}= 6.3\times 10^4\, \rm M_{\odot}$, and $M_{\rm c}^*= 100\, \rm M_{\odot}$. The constants $k_1$ and $k_2$ are obtained from the continuity at $M_{\rm c}^*$ and from the normalization condition $\int_{M_{\rm c,\min}}^{M_{\rm c,\max}} f_{\rm c}(M_{\rm c}) dM_{\rm c} = 1$.
Due to strong stellar light extinction in the Galactic plane, the spatial distribution of SC is known with sufficient accuracy only in the solar neighborhood (for a distance $\lesssim 2$\,kpc from the Sun \cite{Piskunov+2018}). As a consequence $\rho_{\rm c}$ should be derived from some other proxy. Here we use the distribution of pulsars as derived by \cite{Lorimer:2004}, which reads
\begin{equation}
\label{eq:pulsar_dist}
 \rho_c(r) = \left( \frac{r+r_{\odot}}{2 r_{\odot}} \right)^{1.64} 
        \exp{\left[-4.01\,  \frac{r - r_{\odot}}{r_{\odot} +0.55 {\rm kpc}} \right]}
\end{equation}
where $r_{\odot}= 8.5$\,kpc is the Sun position.
On top of the radial distribution, we also account for the distribution inside the Galactic spiral arms, using the same procedure that \cite{Cristofari+2017} adopted to evaluate the distribution of SNRs. The spiral structure is realized by choosing a galactocentric distance $r$ from Eq.\eqref{eq:pulsar_dist} and than by choosing randomly an arm. The polar angle is then determined so that the cluster lies in the centroid of the arm. The actual position of the SC is computed by applying a correction to the galactocentric distance drawn from a normal distribution centered at zero with a standard deviation $0.07\,r$.
Figure~\ref{fig:clusters} shows the result of a single realization of SC population with an age younger than 3 Myr, in terms of radial distribution from the GC and position in the Galactic disk. The total number of clusters results to be $\simeq 300$.

\section{Cosmic ray acceleration}
\label{sec:acceleration}
Following \cite{Morlino+2019}, we assume that the acceleration of particles only occurs at the WTS developed by the SC wind. The (relativistic) energy distribution function of accelerated particles located at the shock position can be written as
\begin{equation}
\label{eq:fs}
 f_s(E) = \xi_{\rm cr} \, \frac{L_{\rm w,c}}{m_p c^2} \frac{1}{\Lambda_p} \left( \frac{E}{m_p c^2} \right)^{-s}  \, e^{-\Gamma(E)} \,,
\end{equation}
where the normalization constant $\Lambda_p$ is defined such that the CR luminosity of the system is equal to $\xi_{\rm cr}$ times the wind luminosity, i.e. $L_{\rm cr} \equiv \int E f_s(E) u_2 \, dE = \xi_{\rm cr} L_{c,w}$, $u_{2}$ being the downstream wind speed.
The solution in Eq.~\eqref{eq:fs} has the typical power-law term $\propto E^{-s}$, found for the case of plane parallel shocks, plus an additional exponential function which accounts for the effects due to the spherical geometry and to the escaping of particles from the bubble boundary, which determine the maximum energy of the system. The exponential function has a complicated expression which, however, can  be approximated by the following formula:
\begin{equation}
\label{eq:Gamma}
 e^{-\Gamma(E)} \simeq \left[ 1+ A (E/E_{\max})^b \right] \, e^{-k (E/E_{\max})^c} \,.
\end{equation}
where $A$, $b$, $k$ an $c$ are fitting parameters, while $E_{\max}$ is the nominal maximum energy defined by the condition that the upstream diffusion length is equal to the shock radius, i.e. $D_1(E_{\max})/u_1=R_s$, $D_1$ being the upstream diffusion coefficient.
The diffusion properties inside the bubble represents the most uncertain parameter of the system. Again following \cite{Morlino+2019} we parameterise the diffusion coefficient as $D= v/3 \, r_L^{\delta} \, \left(r_L/L_c\right)^{1-\delta}$, where $r_L$ is the Larmor radius, while $L_c\simeq 1$\,pc, is the coherence length-scale of the magnetic field, assumed to be of the order of the SC core radius. The exponent $\delta$ is equal to 1/3, 1/2 and 1 for Kolmogorov, Kraichnan and Bohm-like diffusion, respectively. 
Finally, the magnetic field, $\delta B$, used to evaluate the Larmor radius, is estimated assuming that a fraction $\eta_B$ of the wind luminosity at the shock, is converted into magnetic pressure, namely $(\delta B^2/4\pi) \, 4\pi R_s^2 v_w = \eta_B \, \dot{M}\, v_w^2/2$. $\eta_B$ is expected to be of the order of few percent.

On a very general ground, the spectral slope of accelerated particles is determined by the effective compression ratio at the shock, $\sigma$, which includes the velocity of the scattering turbulence. Using hybrid simulations, it has been recently shown \cite{Caprioli+2020} that the downstream turbulence is, in general, more effective than the upstream one in determining the slope, hence here we will include only such an effect. In a parametric form, we can write the mean velocity of the waves downstream as $\bar{v}_{A,2}=\chi_A \, \sqrt{11/8} \, \eta_B^{1/2} \,v_w$, where $v_A$ is the Alfv\'en speed and $\chi_A=0$ for waves that are symmetrically moving in all directions. The comporession ratio is than written as
\begin{equation}
\label{eq:sigma}
 \sigma = \frac{u_1}{u_2 + v_{A,2}}
 = \frac{\sigma}{u_2 + \sqrt{11/8\, \xi_B} \, \chi_A \, u_1} \,.
\end{equation}
The value of the parameter $\chi_A$ obtained from numerical simulations is of the order of few tens of percent. However, in such simulations the magnetic field amplification is only due to the CR streaming, while in the case of stellar winds the magnetic field is more probably determined by MHD instabilities. Hence it is not clear whether the results by \cite{Caprioli+2020} can be straightforwardly applied to our case. As a consequence, we will take $\chi_A$ as a free parameters.

Now we do have all the ingredients to evaluate the CR flux produced by SCs. However, one last caveat need to be addressed. The instantaneous CR luminosity is formally obtained from the particle flux escaping from the wind bubble, which reads $\phi_{\rm esc} = 4\pi R_b^2 \, \left[D \partial f/\partial R\right]_{R=R_b}$, and the total amount of CR injected by a single SC should be given by the integral during its lifetime, $\int \phi_{\rm esc} dT$. However, one can easy realize that such a contribution is always negligible with respect to the amount of particles stored inside the bubble, which is approximately given by $4\pi/3 f_{s} R_b^3$. This apparent inconsistency is given by the fact that the solution provided by \cite{Morlino+2019} is stationary and does not account for the time evolution of the wind-bubble after the adiabatic phase, when the bubble will fade out, releasing all CR stored during the acceleration phase. This issue can be solved replacing the escaping flux with the flux injected inside the bubble, i.e. $\phi_{s} = 4\pi R_s^2 \, u_2 f_s$. \cite{Morlino+2019} have shown that $\phi_{\rm esc}$ and $\phi_s$ differ only slightly at energies $\gtrsim E_{\max}$.

\section{Results and discussion}
\label{sec:results}
\begin{figure}
\centering
\subfigure[\label{fig:chiA0}]{\includegraphics[width=0.49\textwidth]{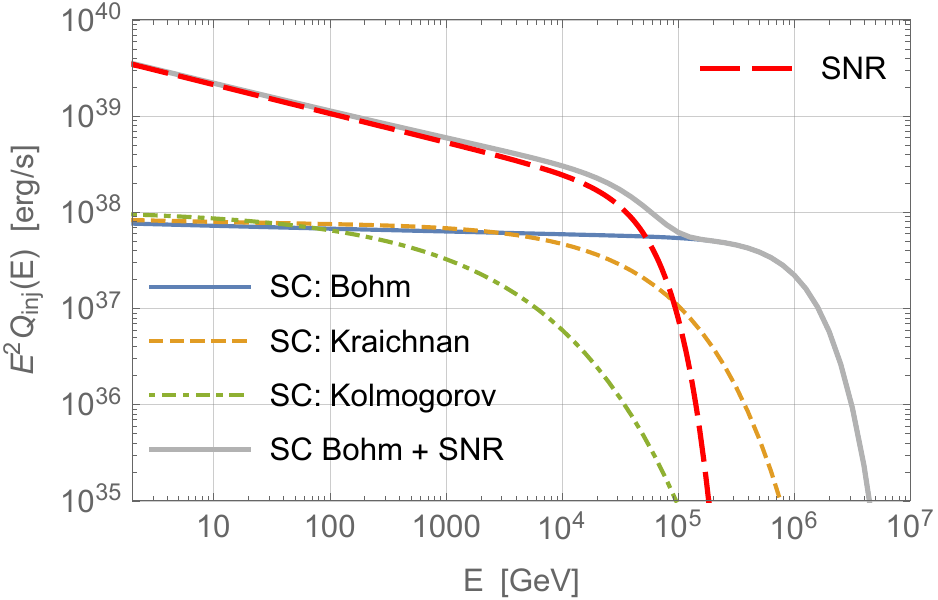}}
\subfigure[\label{fig:chiA01}]{\includegraphics[width=0.49\textwidth]{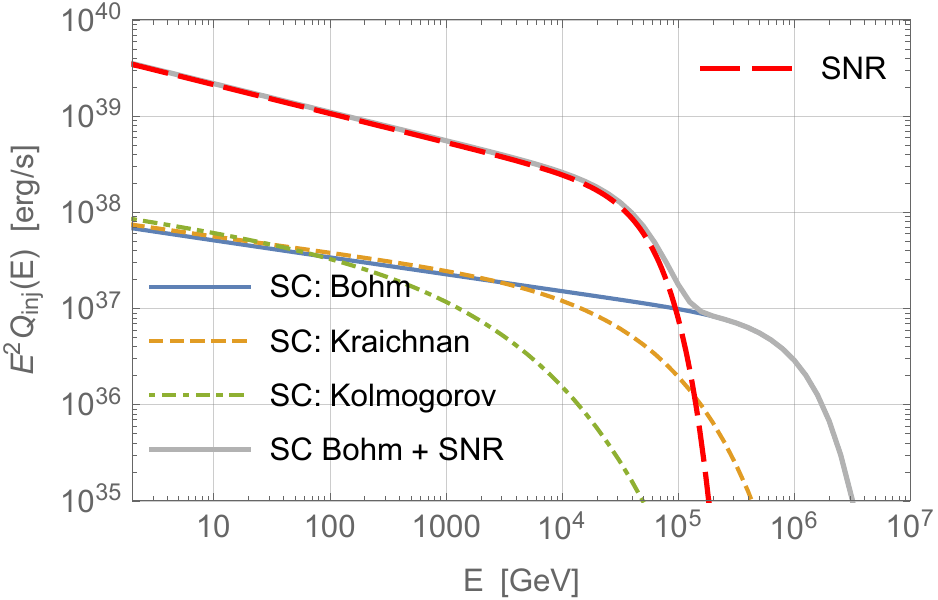}}
\caption{CR spectra injected by stellar clusters, compared with the one injected by SNRs (red-dashed lines). Panels \ref{fig:chiA0} and \ref{fig:chiA01} show the cases corresponding to $\chi_A= 0$ and $\chi_A= 0.1$, respectively. Different lines are calculated with different assumptions for the diffusion coefficient in the wind-bubble, as labelled.}
\label{fig:CR_spectra}
\end{figure}

Using the approximation provided in Eq.\eqref{eq:SC_dist2} we can write the CR flux at the present time $T$ injected by the entire population of SC as follows:
\begin{equation}
\label{eq:Q_SC_2}
 Q_{\rm SC}(E,T) \simeq \bar{\psi} \, \int_{0}^{T} \hspace{-0.2cm}\int_{M_{c,\min}}^{M_{c,\max}}  f_c(M_c,t) \, \phi_{s}(M_c,t)\, dt\, dM  \times \int_{0}^{R_{\rm disk}} \rho_c(r) \, dr \,.
\end{equation}
In the present work, rather then performing the analytical integral, we proceed generating a synthetic population of SC and then we sum up over the contribution of all individual clusters. 
In Figure~\ref{fig:CR_spectra} we show the CR spectrum injected by the synthetic population shown in Figure~\ref{fig:clusters}. Two different cases are considered: $\chi_A=0$ and $\chi_A=0.1$. 
The CR acceleration efficiency and magnetic amplification efficiency are fixed to $\xi_{\rm cr}=0.05$ and $\eta_A=0.05$. The CR spectrum is shown for three different assumption of the diffusion coefficient (Bohm, Kraichnan and Kolmogorov) and is compared with the spectrum of CR injected by SNRs. Notice that all Figures only show the proton CR component. Heavier elements are not discussed here.
One can see that for Bohm diffusion, the $E_{\max}$ reaches PeV energies, and the SC contribution dominates the CR spectrum above $\sim 100$\,TeV. The case with $\chi_A=0$ show a harder spectrum ($s=2.03$) than the one having $\chi_A=0.1$ ($s=2.18$).

Notice that the contribution from SNRs is estimated using some simplifications. We assume that all SN explode into an uniform medium with density 0.1\,cm$^{-3}$. The kinetic explosion energy is fixed to $10^{51}$\,erg and the ejecta mass to $5\, \rm M_{\odot}$. The particle acceleration and escape is than calculated using the model by \cite{Morlino-Celli:2021} which include the magnetic field amplification due to streaming instability (even though in a simplified approach). As one can see from Figure~\ref{fig:CR_spectra}, the effective maximum energy produced by SNRs is only $\simeq 50$\,TeV (similar results are obtained using more refined models like \cite{Cristofari+2020} for the same SNR parameters). 

There are two important aspects to discuss. The relative normalization between the SCs and the SNRe contributions and the different slopes.
The normalization of the CR flux produced by SNRs is obtained taking the Salpeter IMF for the entire Galaxy stellar population and normalizing it to a SFR of $2\,\rm M_{\odot}\, yr^{-1}$ \cite{Elia+2022}. Than we assume that all stars with $M>8\, \rm M_{\odot}$ explode as SNe. This approach gives a rate of $\simeq 1$ SN every century, a factor roughly 2 smaller (but still compatible within the errors) than the one estimated from the combined evidence from external galaxies and from the observation of historical SNe in our Galaxy \cite{Tammann+1994}.
Then, we assume that the SNR acceleration efficiency is the same as the WTS, namely $5\%$. These assumptions gives the power ratio between SCs and SNRs equal to $P_{\rm SC}/P_{\rm SNR}\simeq 8\%$. If one account for the uncertainties of several parameters, such a value ranges between $1\%$ and $30\%$. 
Given the smaller power injected by SCs with respect to SNRs, the possibility that they are responsible for the CR spectrum at PeV energies strongly depends on two quantities: the diffusion coefficient in the bubble and the value of the spectral slope. Concerning the former, only a diffusion close to Bohm-like seems able to produce a maximum energy substantially larger than the one produced by SNRs. In addition, the spectral slope needs to be harder than the one produced by SNR, like the case shown in Figure~\ref{fig:chiA0}. If these two conditions are realized, than SCs could be responsible for the observed CRs at PeV energies. On the other hand, steeper diffusion and/or slope similar (or steeper) than the one produced by SNRs would question the role of SCs, at least in the simplified model presented here.



%
%
%

\end{document}